\documentclass{article}
\usepackage{graphicx}

\usepackage{titlesec}
\usepackage{lipsum}

\usepackage{subcaption}
\usepackage{floatrow}
\usepackage{placeins}

\usepackage[a4paper, margin=4cm]{geometry}

\usepackage{subeqnarray}
\usepackage{bbold}
\usepackage{upgreek}
\usepackage{textgreek}

\usepackage{dcolumn}
\usepackage{amsmath}
\usepackage{bm}
\usepackage{authblk}

\usepackage{xcolor}
\usepackage{color}

\usepackage{hyperref}
\hypersetup{colorlinks=true, urlcolor=blue, citecolor=blue, filecolor=blue, linkcolor=blue}

\title{Creating attosecond breathing ions by coherent shake-up}
\date{}

\author[1]{James Tarrant}
\author[2]{P\v{r}emysl Koloren\v{c}}
\author[3,4]{Margarita Khokhlova}
\author[1,4]{Marco Ruberti}
\author[1]{Vitali Averbukh}

\affil[1]{%
Imperial College London, London SW7 2AZ, UK
}%
\affil[2]{%
Charles University, Prague 116 36, Czech Republic
}%
\affil[3]{%
King's College London, London WC2R 2LS, UK
}%
\affil[4]{%
Max Born Institute, Berlin 12489, Germany
}%

\begin{document}

\maketitle

\section{Abstract}

Shake-up is a fundamental phenomenon in photoionisation of many-electron systems whereby the ionisation of one electron is accompanied by the simultaneous excitation of another. As a single-photon two-electron excitation, it is the most basic manifestation of electron correlation in nature. In a standard experiment using, for example, a synchrotron light source, shake-up states are populated incoherently and their relative excitation intensities describe the process completely. In this work, we use high-accuracy \textit{ab initio} theoretical methods to show that the physics of shake-up differs qualitatively when ionisation is achieved with an attosecond pulse such as those produced by free-electron laser or high-harmonic generation sources. These coherent light sources possess sufficient bandwidth to enable the new phenomenon of coherent shake-up, a process whose result cannot be described solely by the shake-up state populations. We show theoretically that the ensuing coherent many-electron wavepackets exhibit attosecond quantum beats in the form of a radial expansion and contraction of the electron density which we term ``breathing ions''. We predict that such attosecond oscillatory dynamics will persist for as long as nanoseconds before being damped by radiative decay. Our modelling shows that the predicted coherent shake-up dynamics are readily measurable by presently available attosecond-pump attosecond-probe techniques.

\section{Main}

Historically, the process of shake-up~\cite{ShakeUpReview,MOBreakdown} has been studied using monochromatic synchrotron radiation. In 
the atomic case, this leads to emission of photoelectrons with sharply defined energies that are different for each individual shake-up 
state, therefore creating incoherent population of the ionic shake-up series. In recent years, developments in laser generation have 
allowed the creation of pulses with duration as little as a few hundred attoseconds from both high-harmonic generation 
(HHG)~\cite{AttoReview,HHG_Short} and free-electron laser (FEL)~\cite{LCLS_1} sources.
As a result of their short duration, attosecond pulses are intrinsically broad in energy -- a pulse several hundred attoseconds in 
duration will have a bandwidth on the order of 10eV. When used to ionise a system, attosecond pulses therefore associate 
each resulting ionic state not with a single outgoing electronic state but rather with a range of such states with an energy span on the 
same scale as a typical atomic shake-up satellite spectrum. Within the relevant symmetry constraints, this allows different ionic states 
to be associated with overlapping sets of states of the outgoing electron, creating ionic coherence even after this electron has fully 
departed the system. 

In the molecular case, ionic state coherence induced by attosecond photoionisation leads to the charge migration dynamics which have 
been a strong focus of study in attosecond science~\cite{Calegari2014,Isopropanol,CoherenceGlycine}. Such molecular dynamics are fairly 
short-lived, as nuclear geometry distribution arising from zero-point vibrations as well as the nuclear motion subsequent to ionisation rapidly damps the 
coherence and leads to charge localisation over a period of at most a few tens of femtoseconds~\cite{CoherenceGlycine,Decoherence}.
In the atomic case, however, no such coherence damping mechanisms exist, and therefore the electronic coherence of the ionic system can be probed in isolation. 
Recent studies of atomic systems have shown how ionic coherence arises and can be measured during attosecond photoionisation for states split by spin orbit coupling~\cite{SpinOrbitCoherence_Pabst2012,CoherenceProbe_Goulielmakis2010} and other fine structure effects~\cite{HeCoherence_Mehmood2021}, as well as between different one-hole (1h) configurations~\cite{AttoCoherence_Pabst2011}. 
In this work, we focus on valence ionisation below the double 
ionisation threshold, populating an atomic shake-up series which will in general consist of states in which one electron is excited 
into a diffuse orbital. For the higher-energy states in the series, such orbitals can become increasingly Rydberg-like. This hints 
at the expected coherent ionic dynamics: we predict that coherent population of the shake-up series will lead to radial expansion and 
contraction of the multi-electron wavepacket. For the inner-valence ionisation scenario, this is a form of ``frustrated Auger'' 
behaviour~\cite{FrustratedAuger,DecaySuppression_Averbukh2010,QuasicontinuumDecay_Craigie2014}, whereby the filling of an inner-shell vacancy is accompanied by the promotion of a second electron 
into a diffuse orbital slightly short of ionisation, before the inner-shell vacancy subsequently reappears at some later time as the 
diffuse electron returns to a more localised state. These coherent oscillations will persist until the system relaxes via photon 
emission with a typical timescale on the order of nanoseconds~\cite{Ar_Rad_Lifetime_Neu,Ar_Rad_Lifetime_Ion}, 
seven orders of magnitude longer than the duration of the ionising pulse. 

As a representative case study for these dynamics, we focus on the argon atom. The $3s$-ionised state of argon has a strong bound 
satellite series containing almost 50\% 
of the $3s$ hole strength which runs from the main line at 29.2eV up to the double ionisation potential (DIP) at 43.3eV~\cite{Ar_Kr_satellites,Ar_Satellites_Dyall1982}. The energy region containing the $3s^{-1}$ main line and the bound 
satellites can be spanned by an attosecond pulse of duration close to 300as, which is well within the range achievable at XUV to soft 
X-ray frequencies~\cite{APAPS_Vrakking,XUV_PumpProbe,APAPS_XFEL}. Similar satellite spectra exist for ionisation from the outermost 
$s$-orbital in krypton and xenon~\cite{Ar_Kr_satellites}, so it is reasonable to expect that the dynamics in argon will also be 
qualitatively indicative of those resulting from inner-valence ionisation in these heavier noble gases.

For the numerical simulation of this system, we use the \textit{ab initio} technique of 
B-spline restricted correlation space (RCS) algebraic diagrammatic construction 
(ADC) \cite{bsplADC2014,bsplADC_JCTC2018,bsplADC2019,bsplADC_PCCP2019}. This method makes use of B-splines to model both the bound and continuum 
states, allowing accurate treatment of ionisation dynamics. We calculate the ionic states using the ADC(2,2) 
technique~\cite{ADC22}, ensuring that the satellites are treated consistently through the second order of many-body perturbation theory.

After constructing the Hamiltonian, we propagate the system under the action of an ionising laser pulse by numerically solving the time-dependent Schr\"{o}dinger equation, given in atomic units as
\begin{align}
    \hat{H}\Psi(t)=i\frac{\partial}{\partial t}\Psi(t)
\end{align}
\begin{align}
    \hat{H}=\hat{H}_0+\mathcal{E}(t)\cdot\hat{D}-i\hat{W} \, ,
\end{align}
where $\hat{H}_0$ is the RCS-ADC Hamiltonian of the unperturbed system as calculated via the B-spline RCS-ADC method, $\hat{D}$ is the 
corresponding dipole matrix, $\mathcal{E}$ is the applied field vector, and the complex absorbing potential $i\hat{W}$ is imposed on the 
numerical grid boundary to prevent unphysical reflections of the outgoing electron wavepacket. We take the axis of the field 
polarisation to define the z axis. The system is propagated using the Arnoldi-Lanczos algorithm~\cite{ArnoldiLanczos,MarcoArnoldi}.

The resulting ionic system is partially coherent, so, in order to follow the subsequent dynamics, we adopt a density-matrix-based 
approach. Defining the density matrix of the $N$-electron system as 
\begin{align} \label{DensMatDef1}
    \hat{\rho}^N(t)=\left|\Psi^N(t)\right\rangle\left\langle\Psi^N(t)\right| \, ,
\end{align}
where $|\Psi^N(t)\rangle$ is the full $N$-electron wavefunction following the numerical time propagation, we obtain the reduced density 
matrix of the ionic system by taking the trace over the states $|\psi_\mu\rangle$ of the departing electron,
\begin{align} \label{DensMatDef2}
    \hat{\rho}^{N-1}_{\mathrm{red}}(t)=\text{Tr}_\mu\{\hat{\rho}^N(t)\} \, .
\end{align}
The trace runs over the unobserved degrees of freedom, and our construction therefore assumes that no information about the energy or 
emission angle of the emitted electron is retained after it departs the system. We can express the density matrix in terms of the ionic 
eigenstates $|\Psi_n^{N-1}\rangle$ as
\begin{align}
    \hat{\rho}^{N-1}_{\mathrm{red}}(t)=\sum_{m,n}\rho_{m,n}^{N-1}(t)\left|\Psi_m^{N-1}\right\rangle\left\langle\Psi_n^{N-1}\right| \, ,
\end{align}
where the diagonal elements $\rho_{n,n}$ correspond to the populations of the ionic states $n$, while the off-diagonal 
elements $\rho_{m,n}$ with $m\neq n$ indicate the coherence between states $m$ and $n$. In this work we will refer to the fractional 
coherence strengths 
\begin{align}
    G_{mn}(t)=\frac{|\rho^{N-1}_{m,n}(t)|}{\sqrt{\rho^{N-1}_{m,m}(t)\rho^{N-1}_{n,n}(t)}} \, ,
\end{align}
which express each of the coherences as a proportion of their maximum possible value.
The evolution of the density matrix is determined by the von Neumann equation~\cite{Coherences_Ruberti2021}. After the pulse is over, the Hamiltonian is time-independent and the von Neumann equation reduces to
\begin{align}
\rho^{N-1}_{m,n}(t)=e^{-i(E^{N-1}_m-E^{N-1}_n)(t-t_0)}\rho^{N-1}_{m,n}(t_0)
\end{align}
with $E^{N-1}_n$ being the ionic state energies.

The Hartree-Fock basis orbitals used in our calculation are those for the neutral rather than ionic system, so in order to assign configurations to our calculated states we compare to previous numerical and experimental work~\cite{Ar_Kr_satellites,Ar_Satellites_Dyall1982,Ar_satellites_Saloman2010}. In this way, we assign the first few significant satellites of each symmetry to their dominant configurations. These assignments, as well as the energies on which the corresponding Rydberg series converge, are given in the tables below. 

\begin{center}
\begin{tabular}{ |m{3.6cm}|m{3.6cm}|m{3.6cm}|  }
 \hline
 \multicolumn{3}{|c|}{$3s^{-1}$ configurations} \\
 \hline
 Energy / eV & Configuration \{Ne\}... & 1h strength \\
 \hline
  29.06 & ...$3s3p^6$ & 0.5909 \\
  36.33 & ...$3s^23p^4(^1S)4s$ & 0.0093 \\
  38.29 & ...$3s^23p^4(^1D)3d$ & 0.1258 \\
  40.87 & ...$3s^23p^4(^1D)4d$ & 0.0621 \\
  ... & & \\
  44.54 & ...$3s^23p^4(^1D)nd$ & \\
  47.19 & ...$3s^23p^4(^1S)ns$ & \\
 \hline
\end{tabular}
\end{center}

\begin{center}
\begin{tabular}{ |m{3.6cm}|m{3.6cm}|m{3.6cm}|  }
 \hline
 \multicolumn{3}{|c|}{$3p^{-1}$ configurations} \\
 \hline
 Energy / eV & Configuration \{Ne\}... & 1h strength \\
 \hline
  15.86 & ...$3s^23p^5$ & 0.9412 \\
  35.13 & ...$3s^23p^4(^3P)4p$ & 0.0028 \\
  36.76 & ...$3s^23p^4(^1D)4p$ & 0.0135 \\
  38.97 & ...$3s^23p^4(^3P)5p$ & 0.0011 \\
  39.31 & ...$3s^23p^4(^1S)4p$ & 0.0032 \\
  40.67 & ...$3s^23p^4(^1D)5p$ & 0.0021 \\
  ... & & \\
  42.78 & ...$3s^23p^4(^3P)np$ & \\
  44.54 & ...$3s^23p^4(^1D)np$ & \\
 \hline
\end{tabular}
\end{center}

To excite dynamics in this satellite spectrum, we need a pulse which is sufficiently broad to achieve coherent population of the states. 
For this work, we performed initial calculations with a linearly-polarised pump pulse centred at 54.4eV, slightly above the satellite spectrum, with a Gaussian envelope of FWHM 0.358fs in the field amplitude, and corresponding energy FWHM 
10.2eV. This falls within the experimentally accessible range~\cite{APAPS_Vrakking,AttosecondPulse_2006}. The energy width here is sufficient to totally span the $3s^{-1}$ and $3p^{-1}$ satellite spectra excluding the main lines; and is also sufficient to cover the gap from the $3s^{-1}$ main line to the first couple of significant satellites. The gap from the $3s^{-1}$ main line to the satellites just below the DIP, while greater than the energy FWHM, is slightly less than full energy width of the pulse at 20\% of the maximal amplitude, 
so some limited coherence might be expected even between these states. The gap to the $3p^{-1}$ main line is too large to achieve 
coherence to the satellite spectrum with any practical pulse. For our calculations, we chose a pulse intensity of 10$^{13}$W/cm$^2$, 
which falls within the weak-field regime~\cite{Ar_StrongField_Motomura2009} in the sense that the effect of multi-photon and non-resonant transitions is small. 
The pulse carrier envelope phase (CEP) was fixed such that the maximum of the carrier frequency aligns with the peak of the envelope; however, as the pulse duration FWHM is sufficient to contain 5 field cycles, the influence of the CEP on the resulting dynamics is in any case expected to be minor.

\begin{figure}[ht]
\begin{subfigure}{0.49\textwidth}
\includegraphics[width=\columnwidth]{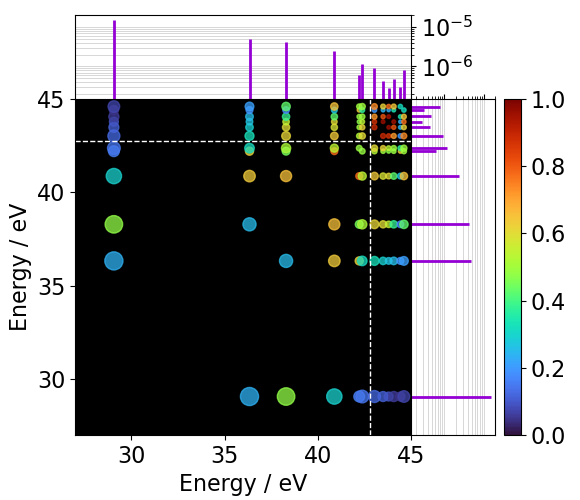}
\caption{}
\end{subfigure}
\begin{subfigure}{0.49\textwidth}
\includegraphics[width=\columnwidth]{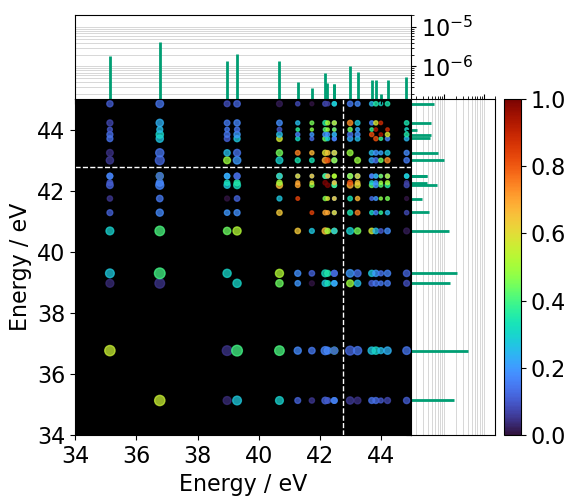}
\caption{}
\end{subfigure}
\caption{The calculated coherence strengths $G_{mn}$ 
of the ionic states up to 45eV following a pulse of central energy 54.4eV and duration FWHM 0.358fs for (a) the $3s^{-1}$ satellite spectrum, and (b) 
the $3p^{-1}_z$ satellite spectrum. The $3p^{-1}$ main line at 16eV has negligible coherence to any of the satellites due to the large energy gap, and is excluded from plot 
(b). The DIP is marked by the dashed white lines. The colour of each circle indicates the coherence strength 
$G_{mn}$, while the size of the circle is proportional to the geometric mean of the corresponding pair of populations such that larger 
circles indicate state pairs which more significantly influence the coherence of the full system. The side panels show the populations 
of each ionic state following the pulse. The $3p^{-1}_{x,y}$ coherences (not shown) are qualitatively similar to the $3p^{-1}_z$ case, but the 
populations are lower especially in the higher-energy states.
}
\label{Ar coherences}
\end{figure}

Our initial calculations found that this pulse produced strong populations of the $3s^{-1}$, $3p^{-1}_z$, and $3p^{-1}_{x,y}$ satellite spectra,
shown for the former two in the side panels of figure~\ref{Ar coherences}. Taken together, the population of the bound $3s^{-1}$ satellite series 
was almost 90\% of that for the main line. For the $3p^{-1}$ satellite series, the satellite series probability was found to be 40\% of the 
main line probability. This is notably higher than the limiting case of sudden ionisation for high pulse energies, where the satellite 
populations would be proportional to the 1h intensity~\cite{SuddenIonisation}. The satellite populations are greater at lower energies, but retain significant magnitude all the way up to the DIP. The discrepancy with the sudden ionisation limit can be understood as arising primarily from the interaction of remaining bound system with the departing electron as it leaves the system over a period of around 1fs.

\subsection{Breathing Dynamics Following Ionisation}

Following the action of the laser pulse, the ionised electron will leave the system, which cannot in general be assumed to be an 
instantaneous process. 
For the pulse that we discuss here, we find that it takes around 4fs for the ionic state populations to stabilise after the pulse, with the large majority of this effect occurring in the first 1-2fs. The change in populations over this period is attributable to residual interaction with the departing electron wavepacket. 
The long-duration dynamics that we are interested in 
are those which occur after this period. The form of these dynamics depends crucially on the remaining coherence in the ionic system. We 
can consider ionic systems corresponding to the $3s^{-1}$ and the different $3p^{-1}$ satellite series separately: coherence between ionic states of 
different symmetry is not possible for a single-photon transition with a linearly polarised pulse as they cannot be associated with the 
same states of the continuum electron. Therefore, since the field we consider is sufficiently weak that the effect of transitions involving multiple 
photons is negligible, the coherence between the different satellite series is essentially zero. Indeed, we find that the density matrix eigenstates can all be assigned to a single symmetry space with a maximal error on the order of $10^{-6}$, which substantiates the assertion that the current pulse intensity can be considered weak. The coherence strengths $G_{mn}$ are 
illustrated for the $3s^{-1}$ and $3p^{-1}_z$ satellite systems in figure \ref{Ar coherences}. For the $3s^{-1}$ spectrum, our results indicate that 
the coherences between the satellites should be fairly strong for the pulse considered, with most of the $G_{mn}$ being in the range 
0.4-0.8. The coherences involving the main line are weaker due to the energy separation, with only the coherence strength to the second 
satellite above 0.4, however even here some coherence is present with states all the way up to the DIP. For the $3p_z^{-1}$ spectrum, 
negligible coherence is expected to the main line due to the energy separation. Amongst the remaining satellites, the degree of 
coherence ranges from near zero to over 0.9, but overall the degree of coherence is substantial.

To assess which states are likely to contribute most to the coherent oscillations, it is useful also to consider the off-diagonal density matrix elements
$\hat{\tilde{\rho}}^{N-1}_{m,n}$ without the previous rescaling, where the ionic density matrix 
$\hat{\tilde{\rho}}^{N-1}_{\mathrm{red}}=\hat{\rho}_{\mathrm{red}}^{N-1}/Tr(\hat{\rho}_{\mathrm{red}}^{N-1})$ is now normalised by the total ionic population. The largest such density matrix elements are given in the tables below for the $3s^{-1}$ and $3p^{-1}_z$ satellite spectra. By this metric, the most significant coherences are those within the $3s^{-1}$ satellite spectrum. The largest is that between the $3s^{-1}$ main line and the \{Ne\}$3s^23p^4(^1D)3d$ satellite, but several other coherences involving the main line are significant due to its comparatively large population, even though the fractional coherence $G_{mn}$ is not especially large. Within the $3p_z^{-1}$ spectrum, the most important coherence by this measure is that between the first two satellites, but the coherences involving various pairs between the first, second, fourth, and fifth satellites are also relevant to the overall coherence of the system. The same is true for the $3p^{-1}_{x,y}$ system. For both the $3s^{-1}$ and $3p^{-1}$ spectra, the most diffuse satellites nearest to the DIP do not contribute so strongly to the overall coherence due to their comparatively lower populations.

\begin{center}
\begin{tabular}{ |m{3.6cm}|m{3.6cm}|m{2.0cm}|m{2.0cm}|  }
 \hline
 \multicolumn{4}{|c|}{$3s^{-1}$ coherences} \\
 \hline
 State $n$ energy / eV (configuration \{Ne\}...) & State $m$ energy / eV (configuration \{Ne\}...) & $\hat{\tilde{\rho}}^{N-1}_{m,n}$ & $G_{mn}$ \\
 \hline
  29.06 (...$3s3p^6$) & 38.29 (...$3s^23p^4(^1D)3d$) & 0.0214 & 0.441 \\
  36.33 (...$3s^23p^4(^1S)4s$) & 40.87 (...$3s^23p^4(^1D)4d$) & 0.0122 & 0.584 \\
  38.29 (...$3s^23p^4(^1D)3d$) & 40.87 (...$3s^23p^4(^1D)4d$) & 0.0118 & 0.603 \\
  29.06 (...$3s3p^6$) & 36.33 (...$3s^23p^4(^1S)4s$) & 0.0113 & 0.217 \\
  29.06 (...$3s3p^6$) & 40.87 (...$3s^23p^4(^1D)4d$) & 0.0101 & 0.272 \\
 \hline
\end{tabular}
\end{center}

\begin{center}
\begin{tabular}{ |m{3.6cm}|m{3.6cm}|m{2.0cm}|m{2.0cm}|  }
 \hline
 \multicolumn{4}{|c|}{$3p^{-1}_z$ coherences} \\
 \hline
 State $n$ energy / eV (configuration \{Ne\}...) & State $m$ energy / eV (configuration \{Ne\}...) & $\hat{\tilde{\rho}}^{N-1}_{m,n}$ & $G_{mn}$ \\
 \hline
  35.13 (...$3s^23p^4(^3P)4p$) & 36.76 (...$3s^23p^4(^1D)4p$) & 0.00834 & 0.497 \\
  36.76 (...$3s^23p^4(^1D)4p$) & 39.29 (...$3s^23p^4(^1S)4p$) & 0.00636 & 0.354 \\
  36.76 (...$3s^23p^4(^1D)4p$) & 40.67 (...$3s^23p^4(^1D)5p$) & 0.00519 & 0.360 \\
  39.29 (...$3s^23p^4(^1S)4p$) & 40.67 (...$3s^23p^4(^1D)5p$) & 0.00478 & 0.464 \\
 \hline
\end{tabular}
\end{center}

In order to illustrate and analyse the electron oscillations, we calculate the hole density as a function of time:
\begin{align}\label{holedenseq}
    Q(\vec{r},t)&=\langle\Psi_0|\hat{R}(\vec{r})|\Psi_0\rangle-\langle\hat{R}(\vec{r})\rangle_{\mathrm{ion}} \nonumber \\[5pt]
    &=\langle\Psi_0|\hat{R}(\vec{r})|\Psi_0\rangle-\text{Tr}\bigg(\hat{R}(\vec{r})\hat{\tilde{\rho}}_{\mathrm{red}}^{N-1}(t)\bigg)
\end{align}
with $\hat{R}(\vec{r})$ the local (electron) density operator. 
$Q(\vec{r},t)$ represents the difference in the electron density at any given point in space between the ionic system and the neutral ground 
state $|\Psi_0\rangle$, so this provides full information about the electron density oscillations.

\begin{figure}[!htb]
\begin{subfigure}{0.34\textwidth}
\includegraphics[width=\columnwidth]{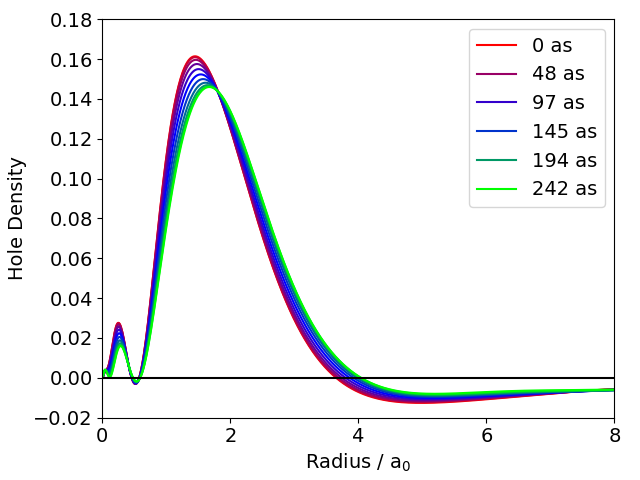}
\caption{}
\end{subfigure}
\begin{subfigure}{0.34\textwidth}
\includegraphics[width=\columnwidth]{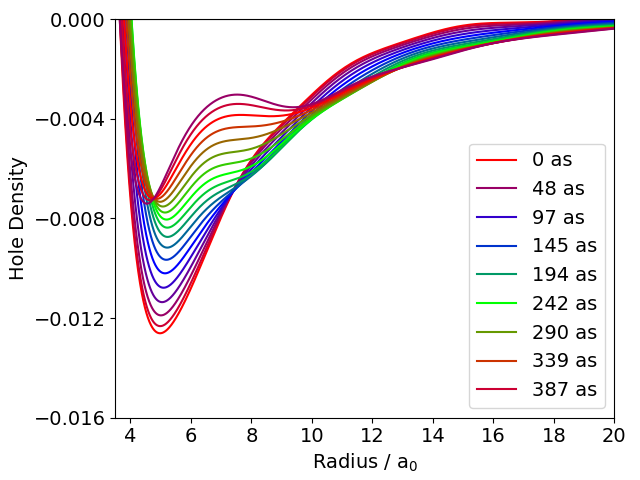}
\caption{}
\end{subfigure}
\begin{subfigure}{0.30\textwidth}
\includegraphics[width=\columnwidth]{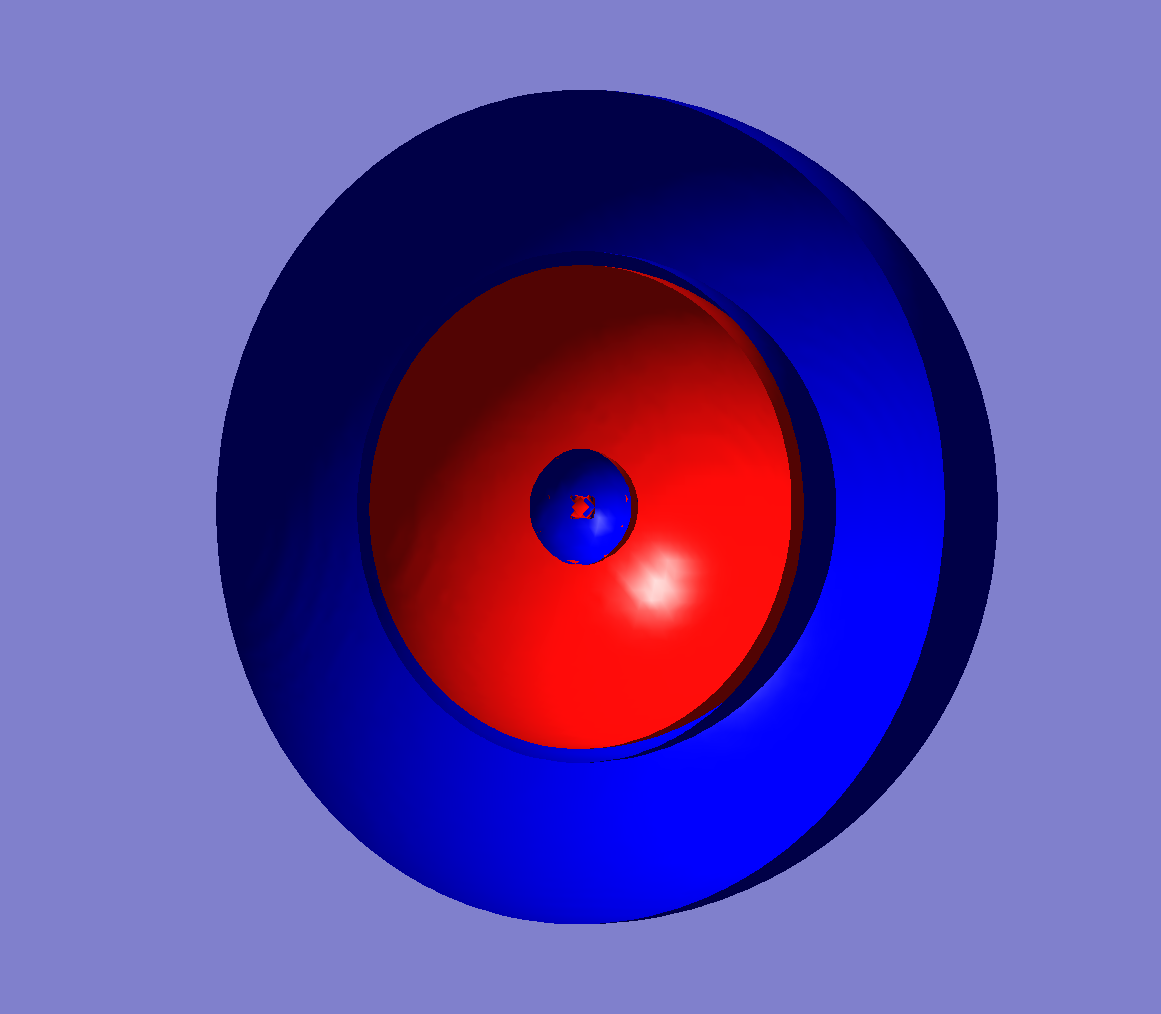}
\caption{}
\end{subfigure}
\begin{subfigure}{0.34\textwidth}
\includegraphics[width=\columnwidth]{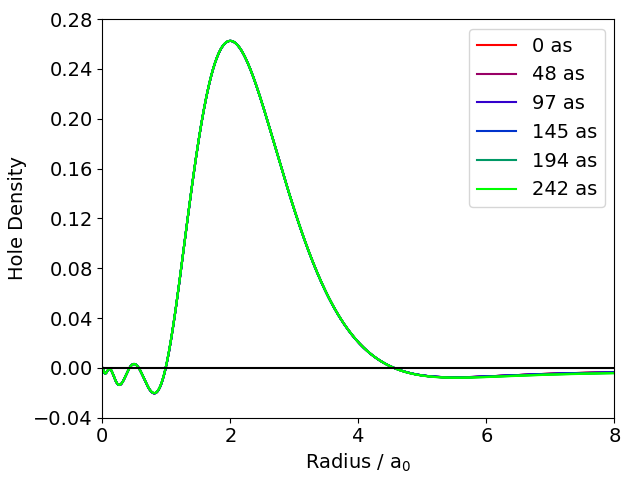}
\end{subfigure}
\begin{subfigure}{0.34\textwidth}
\includegraphics[width=\columnwidth]{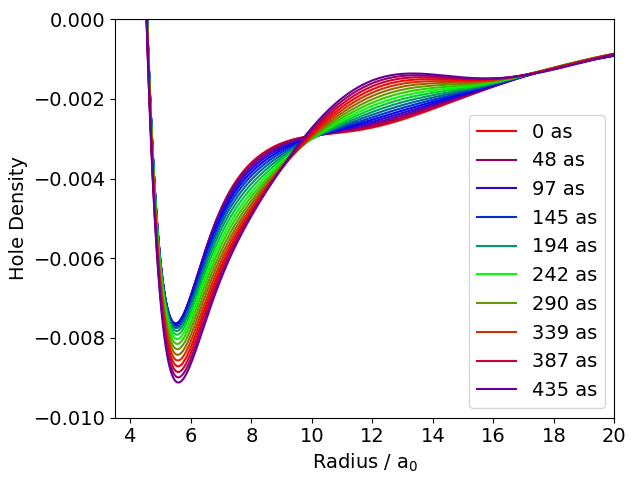}
\end{subfigure}
\begin{subfigure}{0.30\textwidth}
\includegraphics[width=\columnwidth]{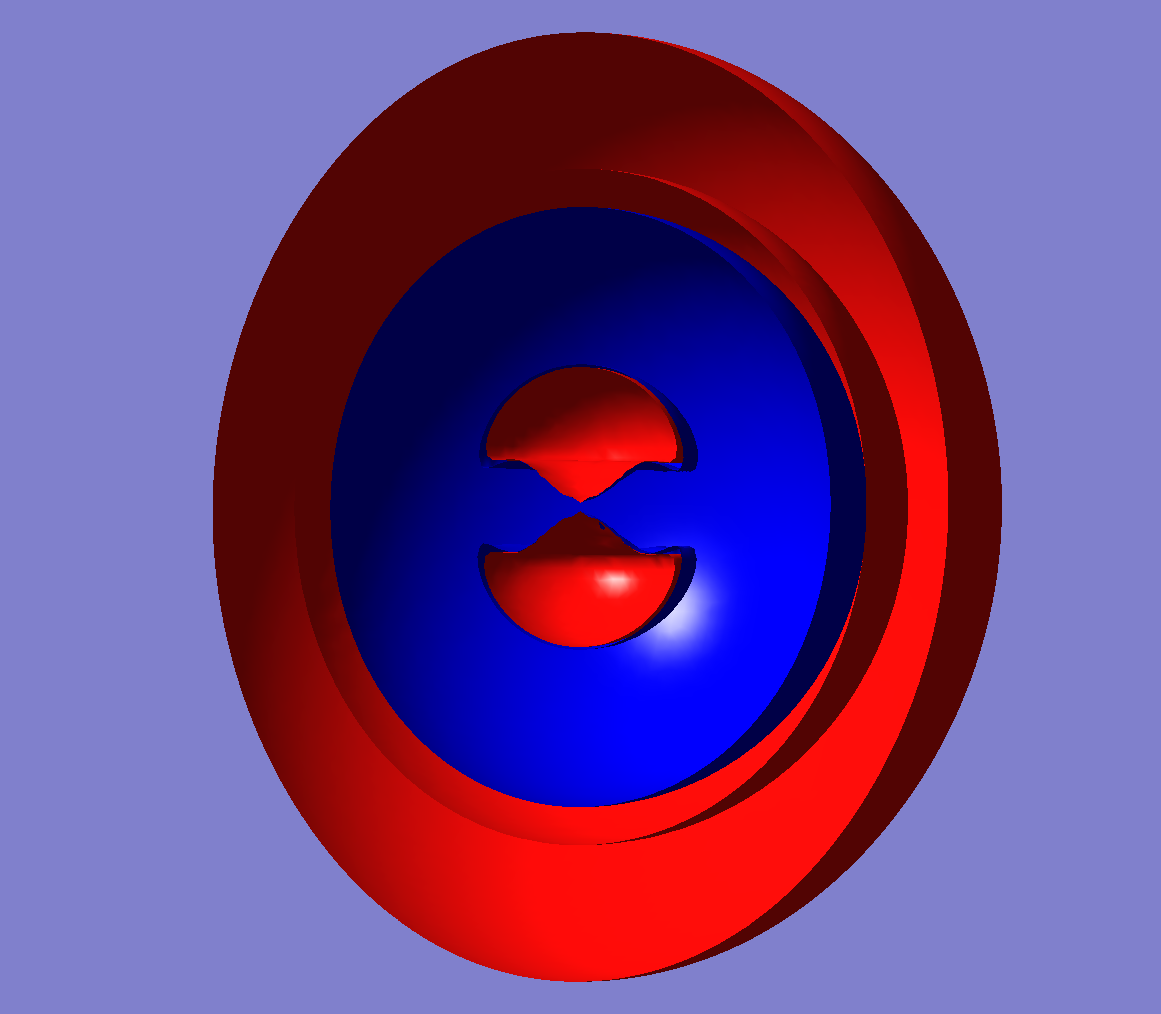}
\end{subfigure}
\caption{For the $3s^{-1}$ satellite system (top) and $3p^{-1}_z$ satellite system (bottom), plots illustrating: (a) the angle-integrated hole 
density as a function of radius up to $8\mathrm{a}_0$, and its variance as a function of time. In this region, the oscillations correspond 
primarily to the 1h part of spectrum. The $3p^{-1}$ contribution is therefore essentially static, while the $3s^{-1}$ peak has a single 
oscillation frequency with a period of approximately 450 as. (b) The angle-integrated hole density as in (a), here focusing on the 
region of negative hole density from 3.5a$_0$ up to 20$\mathrm{a}_0$. The electron motion in this region also includes oscillations between the satellites. The 
motion of the $3s^{-1}$ system in this region is therefore somewhat complex, while the $3p^{-1}$ system oscillation here has one dominant radial 
frequency component. (c) Illustrative density surfaces showing the differences in hole density, $Q(\vec{r},t_2)-Q(\vec{r},t_1)$, across a half-cycle of the most prominent fast radial 
oscillation of each symmetry, up to a radius of $16\mathrm{a}_0$, sliced along a plane parallel to the initial pulse polarisation. The red surfaces indicate 
regions of increased hole density, while the blue regions indicate the opposite. In the $3s$ case, this corresponds primarily to the 
transition from the \{Ne\}$3s3p^6$ state into the \{Ne\}$3s^23p^43d$ states over a time $t_2-t_1$ of 240 as, while for the $3p^{-1}$ system the motion 
is mostly from the \{Ne\}$3s^23p^4(^1D)5p$ into the \{Ne\}$3s^23p^4(^1D)4p$ state and spans a time $t_2-t_1$ of 435 as. The choice of zero time is 
made separately and arbitrarily for the $3s^{-1}$ and $3p^{-1}$ spectra to best illustrate the range of the electron motion.}
\label{rdens figs}
\end{figure}

We now restrict our consideration to the set of ionic states below the DIP. The Rydberg-like states above the DIP are expected to decay on a timescale from a few tens to hundreds of femtoseconds~\cite{RydbergDecay_Gokhberg2007} and will therefore produce transient oscillations which, while themselves moderately long lived, will not significantly influence the majority of the nanosecond-duration dynamics which characterise this system. As we expect the primary component of the oscillations to be radial, we calculate the total angle-integrated hole density as a function of radius, illustrated in figure \ref{rdens figs}. Above a radius of approximately 4a$_0$ - corresponding roughly to the radius of the neutral atom - the hole density falls below zero and remains negative all the way up to 20a$_0$. This region of excess electron density corresponds to the population of satellites in which an electron is excited into a diffuse orbital. As a result of the coherence within the satellite spectrum, the density exhibits clear dynamics.

\subsection{Probes of the Breathing Dynamics}

Up to this point, we have demonstrated that ionisation of argon by an attosecond laser pulse can be expected to induce oscillatory 
electron dynamics. The dynamics that we predict contain both fast oscillations with a period of as little as 450 attoseconds, as well as slower 
contributions with periods of several femtoseconds. For such a system, the most suitable measurement technique is attosecond pump-attosecond probe spectroscopy (APAPS)~\cite{LCLS_1,Isopropanol,APAPS_XFEL}. To detect the faster oscillation here would require a probe 
of several hundred attoseconds in duration, as well as knowledge of the pump-probe delay down to a few tens of attoseconds, both of 
which are within the achievable range~\cite{APAPS_Vrakking}.

We numerically simulate an attosecond probe of the system for a range of probe delays and calculate the resulting probability of double 
ionisation as a function of this delay. The delay is relative to an essentially arbitrary offset, chosen here to be 4.5 fs, which must be 
long enough for the transient oscillations to have decayed and for the electrons ionised by the pulse to have departed the system. We 
simulate two different sets of probe pulse parameters, both polarised parallel to the pump pulse with a Gaussian envelope. The first probe configuration is modelled on one recently
achieved in a practical HHG-based APAPS experiment~\cite{APAPS_Vrakking}, with central energy 33eV and duration FWHM of 253 as (energy FWHM 14.4 eV). This 
pulse energy is sufficient to directly ionise any state in the ionic spectrum, and the pulse duration is expected to be sufficiently 
short to detect the fastest predicted dynamics. The second probe we model is of the single-photon laser-enabled Auger decay (spLEAD) type~\cite{spLEAD} for the $3s^{-1}$ main line. This probe has duration FWHM 376 as (energy FWHM 9.7 eV) and a central energy of 22 eV, which is sufficient for ionisation of the $3s$ hole main state only if the $3s$ hole is simultaneously filled by a $3p$ electron. This process is allowed only after accounting for configuration mixing, making this a direct measure of the 2h1p part of the system. 
We again take the CEP such that the maxima of the envelope and the carrier are aligned, and we choose a pulse intensity of 
10$^{14}$W/cm$^2$. We find that this pulse still falls at least approximately within the regime in which the ionisation rates rise linearly with the intensity. 
The propagation is carried out in an analogous manner to the numerical propagation of the neutral, calculating the ionic Hamiltonian and 
dipole matrix using the equivalent B-spline RCS-ADC method for the ionic system.

\begin{figure}[!ht]
\begin{subfigure}{0.48\textwidth}
\includegraphics[width=\columnwidth]{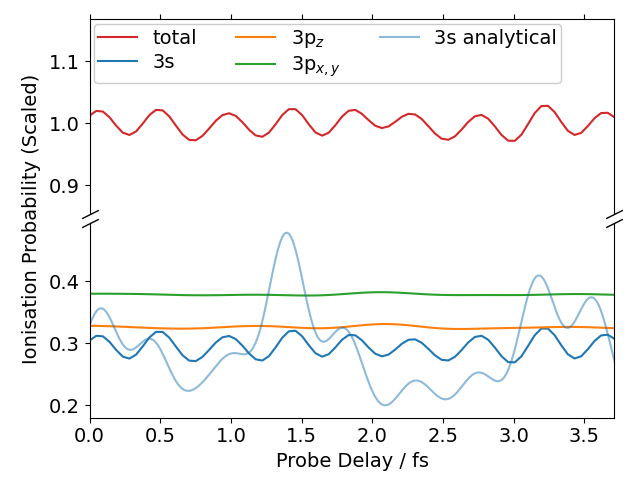}
\caption{}
\end{subfigure}
\begin{subfigure}{0.48\textwidth}
\includegraphics[width=\columnwidth]{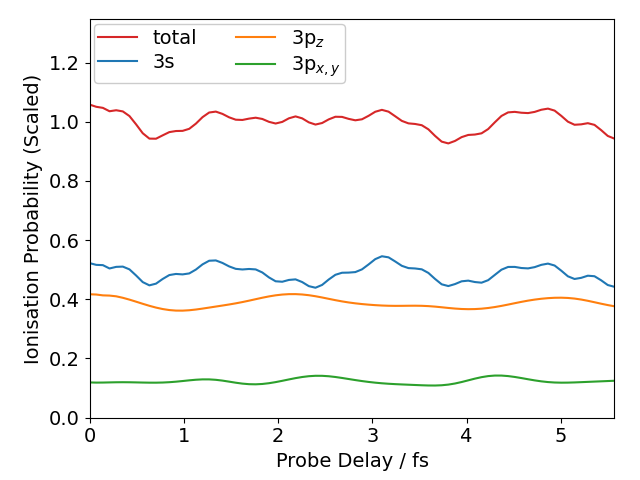}
\caption{}
\end{subfigure}
\begin{subfigure}{0.48\textwidth}
\includegraphics[width=\columnwidth]{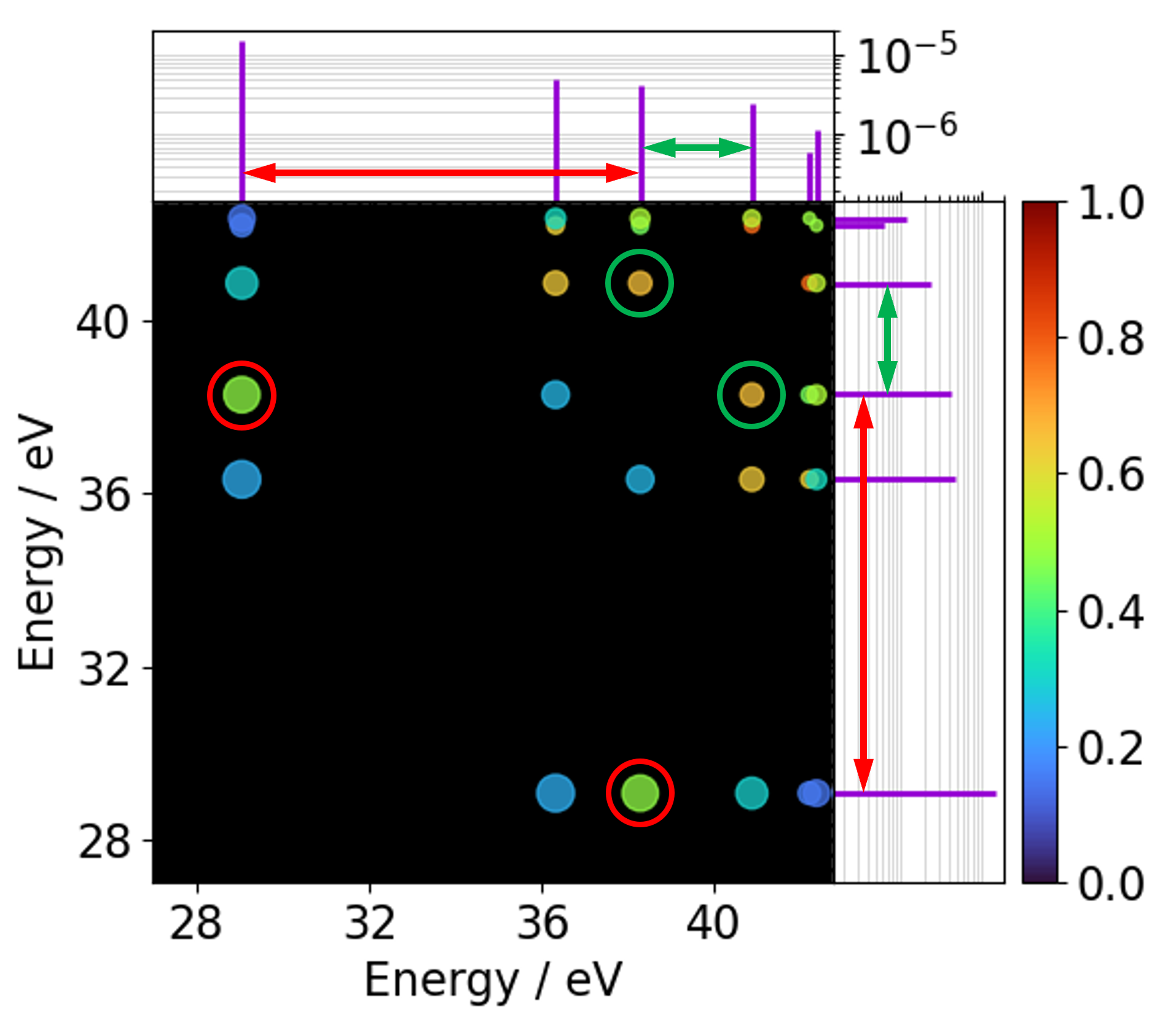}
\caption{}
\end{subfigure}
\caption{The variation of the total double ionisation probability with probe delay relative to an arbitrary offset long after the initial 
pump (red), as well as the contributions from the satellite spectra of each symmetry, 
for (a) a probe of central energy 33 eV and duration FWHM 253 as; and (b) a probe of central energy 22 eV and duration FWHM 
379 as; both scaled such that the average total ionisation probability is 1. Panel (a) also shows the semi-quantitative analytical prediction of the probe ionisation rate for the $3s^{-1}$ spectrum, scaled by an arbitrary factor to align with the \textit{ab initio} results. Panel (c) again illustrates the coherence strengths $G_{mn}$ within the bound $3s^{-1}$ satellite series, now highlighting those coherences which can be detected by the probe. Two coherences in this system can be associated with significant detectable 
oscillations: the coherence between the $3s^{-1}$ main line and the \{Ne\}$3s^23p^43d$ satellite (red circles, arrows) is clearly detectable 
for both probes as a fast oscillation with period 0.45 fs; while the 22 eV probe also clearly picks up a second, slower oscillation with 
period approximately 1.65 fs corresponding to the coherence between the \{Ne\}$3s^23p^43d$ and \{Ne\}$3s^23p^44d$ satellites (green 
circles, arrows).
}
\label{ProbePulse}
\end{figure}

Figure \ref{ProbePulse} illustrates the calculated dependence on the pump-probe delay of the probability that the system becomes doubly ionised by the probe. The 33 eV probe can be seen to primarily detect the oscillation between the $3s^{-1}$ main line and the second satellite, providing a 
measure of the oscillation of the $3s$ hole occupation. The 22 eV probe also detects this coherence, albeit not as strongly since the 
pulse FWHM is on a similar order to the oscillation period, but it also picks up a second frequency corresponding to the coherence 
between the second and third $3s^{-1}$ satellites. Some oscillations corresponding to the $3p^{-1}$ satellite system also appear for the 22 eV 
pulse with a period of around 2.5 fs, likely corresponding to the coherence between the first two $3p_z^{-1}$ satellites, however the prominent oscillation of period approximately 870 as visible in the bottom panels of figure \ref{rdens figs} is not detected here. For the 33 eV pulse, the delay-independent ionisation from the $3p^{-1}$ main line makes up the large majority of the contribution from the $3p^{-1}$ satellite system and a significant fraction of the double ionisation probability overall; however this is significantly suppressed in the case of the 22 eV pulse as only the pulse tail has sufficient energy to ionise this state. 
The 33 eV numerical results can be compared to the results of our semi-quantitative analytical calculation (details in supplementary material). The oscillation corresponding to the coherence between the main line and second satellite appears similarly in both the analytical model and the \textit{ab initio} calculation; however it is clear that the full numerical treatment is required to accurately treat the oscillations arising from coherence between pairs of satellites. 
Our numerical results overall indicate that the electron wavepacket oscillations predicted in the previous section can be observed in the delay-dependent double ionisation rate, with distinctive frequencies corresponding to particular coherently-populated ionic state pairs.

To summarise, our work predicts from first principles that currently available attosecond pulses are capable of inducing strong coherence with atomic systems, and that this coherence manifests in the form of long-duration oscillatory dynamics. Furthermore, our work suggests specific pulse parameters for which such oscillations should be clearly detectable using an experimentally-realisable attosecond pump-attosecond probe scheme in the case of the argon atom. We illustrate that the frequency components of the detected oscillations provide an indication of the most prominent coherences in the system.

\clearpage

\section{Methods}

As noted previously, the numerical calculations in this work make use of the B-spline RCS-ADC method~\cite{bsplADC2014,bsplADC_JCTC2018,bsplADC2019,bsplADC_PCCP2019} at the ADC(2)x level of approximation~\cite{MBMSchirmer}. This scheme divides the basis into two orthogonal spaces, within each of which the Fock matrix is diagonal: one, the restricted correlation space (RCS), contains all contributions $\chi_\alpha$ to the (exact) neutral ground state; while the other, the ionisation space (IS), is defined such that the annihilation operator for any orbital $\psi_\mu$ in this space annihilates the ground state. The IS contains all of the continuum-like orbitals associated with the outgoing electron after ionisation. The primary reason for this choice of construction is that it massively reduces the computational cost of the ensuing ADC calculation; however, an additional effect is that in the final construction of the RCS-ADC Hamiltonian in the intermediate state representation, the intermediate state basis corresponding to a particle in the IS decomposes as 

\begin{align}
    |\Psi^N_{\mu,I_{RCS}}\rangle=c^\dagger_\mu|\Psi^{N-1}_{I_{RCS}}\rangle \, ,
\end{align}

\noindent where $I_{RCS}$ corresponds to a 1h or 2h1p configuration of RCS states, $c^\dagger_\mu$ is a creation operator corresponding to the state $\psi_\mu$, and $|\Psi^{N-1}_{I_{RCS}}\rangle$ is an intermediate state of the $(N-1)$-electron system. Subject to a rotation of basis, this allows the part of the Hamiltonian basis containing the IS to be expressed in terms of the ionic eigenstates $|\Psi^{N-1}_n\rangle$ as

\begin{align} \label{RCSIonicSplit}
    |\Psi^N_{\mu,n}\rangle=c^\dagger_\mu|\Psi^{N-1}_n\rangle \, ,
\end{align}

\noindent where the ionic states are calculated only in the smaller RCS basis. The expected dynamics of the ionic system are driven by the 2h1p-dominated satellite states, so it is necessary to model these accurately. Within the ADC(2)x level of theory, 1h-dominated ionic states are treated consistently through the second order of perturbation theory, but states consisting primarily of 2h1p configurations are treated only at first order. This is insufficient to accurately capture the form of these satellites, so we improve our description by calculating the ionic states separately at the ADC(2,2) level of theory~\cite{ADC22}, which treats both 1h- and 2h1p-dominated ionic states consistently through the second order of perturbation theory. As this level of treatment requires the inclusion of 3h2p configurations, these states cannot be incorporated directly, and this 3h2p component must be first stripped away. We retain only those ionic states which have more than 1\% 1h character, as the population of the remaining states by a short ionising pulse in the weak-field regime can be assumed to be negligible. For these states, the deleted 3h2p configurations typically account for around 10\% of the total makeup. These states are then symmetrically re-orthonormalised, and taken to replace the ionic states in the construction of the Hamiltonian as in \eqref{RCSIonicSplit}.

As we expect the ionic system to have highly diffuse oscillations, we must model the ionic states using a basis which is capable of representing such diffuse orbitals. For this purpose, we augment a core GTO basis of the ANO-RCC-VQZP type~\cite{ANO-RCC-VQZP} with a set of diffuse basis functions $\zeta_{n,l}=r^l \text{exp}(-\alpha_{n,l} r^2)$ where the coefficients $\alpha$ are optimised to represent rydberg-like orbitals~\cite{KBJ_continuum} with effective charge $Z=1.5$. For the diffuse part of the basis, we use 12 functions with angular momentum $l=0$, 12 with $l=1$, 12 with $l=2$, 10 with $l=3$, and 3 with $l=4$.

The density matrix obtained after the calculation as in \eqref{DensMatDef2} must be corrected to account for the fact that the departing electron wavefunction has been partially absorbed by the complex absorbing potential. Defining the integral $\langle\Psi^N(t)|c_\mu^\dagger|\Psi_n^{N-1}\rangle=a_{n\mu}(t)$, this correction takes the form~\cite{bsplADC_PCCP2019}

\begin{align}
\rho^{N-1}_{m,n}(t)=\sum_\mu a_{n\mu}(t)a_{m\mu}^*(t)+2\int_{-\infty}^t dt' \sum_{\mu\nu} a_{n\nu}(t')a_{m\mu}^*(t') w_{\nu,\mu} e^{i(I_n-I_m)(t-t')}
\end{align}

where the $I_n$ are the ionisation potentials of ionic state $n$, and the $w_{\nu,\mu}$ represent the matrix elements $\langle\nu|\hat{W}|\mu\rangle$ of the CAP between the ionisation space orbitals $\nu$ and $\mu$.

To obtain the hole density as a function of the spatial coordinates as in equation \eqref{holedenseq}, we express the local electron density operator in terms of the Hartree-Fock orbitals $\phi_i(\vec{r})$ used to build the RCS

\begin{align}
    \hat{R}(\vec{r})=\sum_{i,j}\phi^\ast_i(\vec{r})\phi_j(\vec{r})c^\dagger_ic_j
\end{align}

The hole density can be re-expressed in a diagonal form~\cite{bsplADC_PCCP2019}

\begin{align}\label{NaturalChargeOrbitals}
    Q(\vec{r},t)=\sum_{i}|\tilde{\phi}_i(\vec{r},t)|^2 \tilde{n}_{i}(t)
\end{align}

in terms of the now both space- and time-dependent natural charge orbitals $\tilde{\phi}_i(\vec{r},t)$ and their populations $\tilde{n}_{i}(t)$. 

The partially-coherent density matrix which represents the system following ionisation can be diagonalised to re-express the system in terms of a series of totally mutually incoherent states $|\xi_j^{N-1}(t)\rangle$, with the eigenvalues corresponding to the state populations $p_j(t)$
\begin{align}
    \hat{\rho}^{N-1}_{\mathrm{red}}(t)=\sum_{j}p_j(t)\left|\xi_j^{N-1}(t)\right\rangle\left\langle\xi_j^{N-1}(t)\right|
\end{align}
Expressing the ionic density matrix as such a sum over mutually incoherent states corresponds to a Schmidt decomposition of the neutral system, and is known as purification of the reduced density matrix~\cite{Coherences_Ruberti2021}. We use this approach to calculate the action of the probe pulse: we propagate each density matrix eigenstate individually under the action of the Schr\"{o}dinger equation, before recombining the results in an incoherent sum weighted by the populations. Each coherent state is constructed from a sum of density matrix eigenstates in the RCS basis, and the propagation is carried out analytically up to the start of the probe pulse without including the continuum states. During the probe, we switch to numerical propagation within the full RCS + IS basis, and we then project the final wavefunction of the system onto the eigenstates of the ADC(2)x Hamiltonian, obtaining the ionisation rate as the probability that the system finishes in a state above the DIP.

\FloatBarrier

\section{Acknowledgements}

MR and VA acknowledge support from the EPSRC grant “Quantum entanglement in attosecond ionization,” Grant No. EP/V009192/1. 
MK acknowledges Royal Society funding under URF\textbackslash R1\textbackslash 231460.

\section{Contributions}

MR and VA jointly designed and supervised the work. JT performed the first-principles many-electron simulations. JT performed the interpretation of the first-principles simulations results with contributions from all authors. PK contributed to and supported the many-electron simulations. MK developed the analytical model. The manuscript was written by JT with contributions from all authors.

\clearpage

\section{Extended Figures}

\begin{figure}[!htb]
\begin{subfigure}{0.49\textwidth}
\includegraphics[width=\columnwidth]{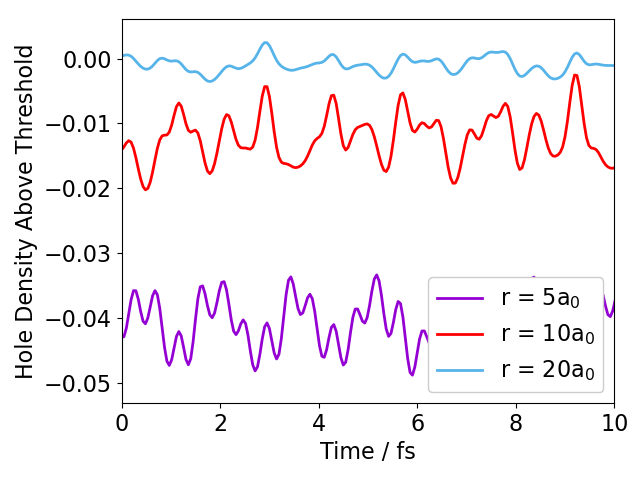}
\end{subfigure}
\begin{subfigure}{0.49\textwidth}
\includegraphics[width=\columnwidth]{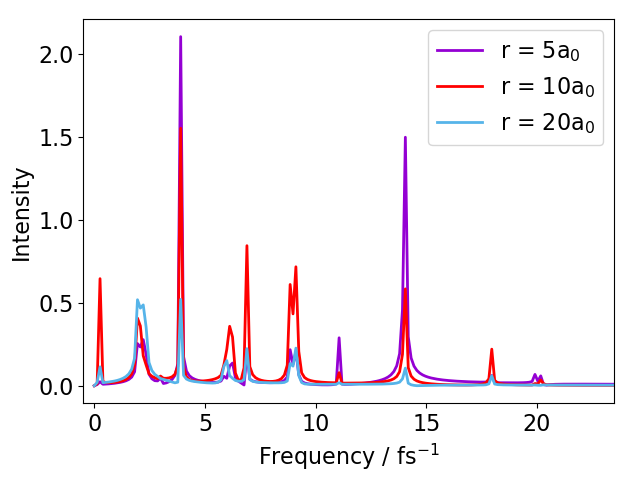}
\end{subfigure}
\begin{subfigure}{0.49\textwidth}
\includegraphics[width=\columnwidth]{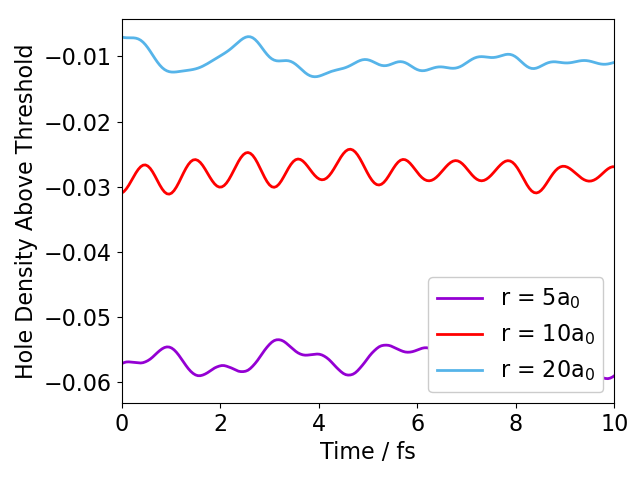}
\end{subfigure}
\begin{subfigure}{0.49\textwidth}
\includegraphics[width=\columnwidth]{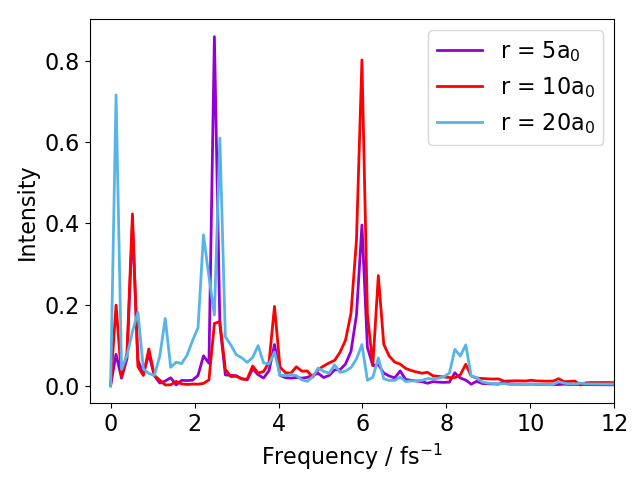}
\end{subfigure}
\caption{The total integrated hole density above a series of radial thresholds (left), and the frequency components of these oscillations (right). The top pair of plots corresponds to the electron density associated with the $3s^{-1}$ system, while the bottom pair corresponds to the $3p^{-1}$ system. As these oscillations are long-lived and do not change their form once the ionised electron has left the system, the choice of zero time is essentially arbitrary and is taken here to be 4.5 fs after the peak of the ionising pulse. It can be seen that a small number of coherences involving lower energy satellites dominate the dynamics just above the radius of the neutral atom, whereas at larger radii a greater number of coherences influence the dynamics, especially for the $3s^{-1}$ system. The fast dynamics illustrated in figure \ref{rdens figs} correspond primarily to the $3s^{-1}$ oscillations with frequency 14 fs$^{-1}$ and to the $3p^{-1}$ oscillations at 6fs$^{-1}$ seen here; while the dynamics probed in figure \ref{ProbePulse} closely match the 5a$_0$ radius dynamics of the $3s^{-1}$ system, especially for the 22eV probe.}
\label{Diffusedensity}
\end{figure}

\clearpage

\bibliographystyle{ieeetr}

\bibliography{bibliography}

\end{document}